\documentclass[]{aastex}          
\usepackage{spr-astr-addons}    

\begin{document}
%
\title{Searching for star-planet magnetic interaction in CoRoT observations$^{*}$}

\shorttitle{Star-planet magnetic interactions}
\shortauthors{A.~F.~Lanza}

\author{A.~F.~Lanza\altaffilmark{1}} 

\altaffiltext{1}{INAF-Osservatorio Astrofisico di Catania, Via S.~Sofia, 78 - 95123 Catania, Italy; e-mail:nuccio.lanza@oact.inaf.it \\
~\\
$^{*}$Based on observations obtained with CoRoT, a space project operated by the French Space Agency, CNES, with partecipation of the Science Program of ESA, ESTEC/RSSD, Austria, Belgium, Brazil, Germany, and Spain.}

\begin{abstract}
Close-in massive planets interact with their host stars through tidal and magnetic mechanisms. In this paper, we {  review} circumstantial evidence for star-planet interaction as revealed by the photospheric magnetic activity in some of the CoRoT  planet-hosting stars, notably CoRoT-2, 4, and 6. The phenomena are  discussed in the general framework of  activity-induced features in stars accompanied by hot Jupiters. 
The theoretical mechanisms proposed to explain the activity enhancements possibly related with hot Jupiters are also  briefly reviewed with an  emphasis on the possible effects at the photospheric level. The unique advantages  of CoRoT and Kepler observations to test these models  are pointed out.  
\end{abstract}

\keywords{stars: planetary systems -- stars: activity -- stars: late-type -- stars: magnetic fields -- stars: general}

\section{Introduction}
Massive planets in close orbits {  \citep[semimajor axis $ a\leq 0.15$~AU; see, e.g., ][ Sect.~3.3 for a motivation for this choice]{Kashyapetal08} } around late-type stars can interact with their hosts through tidal as well as magnetic mechanisms \citep[see, e.g., the pioneering work by ][]{Cuntzetal00}. 
Although the study of the tidal interaction is of fundamental importance to understand the formation and evolution of these systems \citep[see, e.g., ][]{Dobbs-Dixonetal04,PapaloizouTerquem06,GoupilZahn08,Zahn08,Lanzaetal11a}, here we shall focus on the interaction involving magnetic fields and the associated activity features. The first evidence for such an interaction was reported by \citet{Shkolniketal03} and later by \citet{Shkolniketal05,Shkolniketal08}.  We shall {  review} the phenomena observed in some of the hosts of the CoRoT planets that may suggest some kind of magnetic connection between the stars and their hot Jupiters. Since the observations are not conclusive yet, further studies are recommended to confirm and clarify the interaction. We  stress the possibilities offered by high-precision space-borne photometers, like CoRoT and Kepler, for this kind of studies and briefly review the mechanisms suggested to explain how a close-in massive planet may affect the activity of its host star. 

\section{Observations of star-planet magnetic interactions}

\citet{Shkolniketal03} and \citet{Shkolniketal05} observed a sample of {  twelve} hot Jupiter systems reporting some evidence for a stellar activity modulated with the orbital period of the planet rather than the star rotation period. The cases of HD~179949 and $\upsilon$~And are those displaying the clearest effects {  in observations of modulated Ca~II H\&K chromospheric emission.}  Their chromospheric hot spots rotating in phase with the planet irradiate powers up to $\sim 10^{21}$~W and are not located at the subplanetary longitude, but lead the planet by $\approx 80^{\circ}$ in the case of HD~179949 and  $\approx 160^{\circ}$ in the case of $\upsilon$~And. In both the cases, the hot spots seem to have a lifetime of $\sim 300-400$~days and are detected  only in {  $\sim 50-60$ percent of the seasons.} Therefore, \citet{Shkolniketal08} suggested that the star-planet magnetic interaction (hereafter SPMI) undergoes transitions from "on" to "off" states and viceversa. Observations by \citet{Poppenhaegeretal11} did not detect any hot spot modulated in phase with the planet in  $\upsilon$~And during their campaign in 2009. 

The signatures of SPMI were searched also in the coronal X-ray emission. \citet{Kashyapetal08} claimed that stars with hot Jupiters ($a < 0.15 $~AU) have a higher X-ray flux than stars with distant ($a > 1.5$~AU) planets, with an enhancement by a factor of $\approx 2$. However, \citet{Poppenhaegeretal10} cast  doubts on this result and found no evidence of an X-ray emission enhancement by comparing  volume-limited samples of stars with and without hot Jupiters. They explain  the result of \citet{Kashyapetal08} as a consequence of the selection effects affecting their samples. 
Claims by \citet{Scharf10} and \citet{Hartman10} about a relationship between stellar X-ray luminosity and the mass or the surface gravity of the planet, respectively, were also refuted as consequences of subtle selection effects affecting the samples used by those investigators \citep[see ][ and the contribution by Katja Poppenhaeger to the 2nd CoRoT Symposium]{PoppenhaegerSchmitt11}. However, an interesting observation was reported by \citet{Pillitterietal10} who found a softening of the X-ray spectrum  and a strong flare close to the secondary eclipse of the hot Jupiter orbiting HD~189733. Although this could simply be a coincidence, a coronal flaring activity modulated with the orbit of the planet cannot be excluded. 

Interesting results came from observations in the ultraviolet, performed mainly with the Hubble Space Telescope.
The first studies \citep[e.g., ][]{Vidal-Madjaretal03,Ehrenreichetal08} found an extended hydrogen envelope around the transiting planet HD~209458b likely to be the result of the evaporation of the planetary atmosphere under the action of the stellar irradiation. More recent studies \citep[][]{Fossatietal10} detected an excess absorption in the near UV during transit ingress in WASP-12.  \citet{Llamaetal11} modelled it as resulting from the absorption of the stellar coronal plasma compressed at a bow shock formed at the boundary with the planetary magnetosphere. Although the present observations are still preliminary, UV transit light curves can offer a unique opportunity to detect such bow shocks, as theoretically discussed by, e.g., \citet{Vidottoetal11}. The shock characteristics depend on the geometry and intensity of the planetary magnetic field as well as on the properties of the stellar coronal plasma and the stellar wind. Therefore, they can be used to derive information on the planetary magnetic fields given that the stellar coronal field can be modelled through the extrapolation of  the photospheric field measured by spectropolarimetric techniques \citep[e.g. ][]{Moutouetal07,Faresetal10}.

SPMI signatures in the photospheres of planet-hosting stars have been suggested in the case of $\tau$~Bootis and HD192263. \citet{Walkeretal08} observed $\tau$~Boo for five years both from the ground by monitoring the flux in the core of the chromospheric Ca~II~H\&K lines and from  the space with the MOST (Microvariability and Oscillation of Stars) microsatellite. A persistent active region was found on the star at a longitude leading the subplanetary longitude by $\approx 80^{\circ}$ with a variable brightness and a lifetime of at least $\approx 1400$~days. 
Since the rotation of $\tau$~Boo is almost synchronized with the orbital motion of its planet, it was the persistence of the active region that pointed towards a SPMI effect. 

In the case of HD~192263, \citet{Santosetal03} reported evidence of starspots rotating with the planet orbital period for several rotations and lagging the subplanetary longitude by $\approx 90^{\circ}$. This may produce a confusion between the radial velocity jitter induced by activity and the actual reflex motion of the star produced by the planet since both have the same period, but the former lacks the long-term phase coherence that characterises the latter. Further evidence of photospheric features possibly associated with SPMI has been extracted from CoRoT photometry, as we discuss in the next section. 

For completeness, we mention also the possibility of detecting SPMI in the radio domain. The method is promising {  because it can allow us to directly measure the planetary magnetic field (cf. end of Sect.~\ref{chromo_features}), }  although no confirmed radio detections of exoplanets have been reported to date \citep{HessZarka11,Lecavelierdesetangsetal11}. 

\section{Photospheric SPMI from CoRoT observations}
\label{phot_SPMI_corot}

We  introduce three cases, i.e., those of CoRoT-2, CoRoT-4, and CoRoT-6. In the case of CoRoT-2, the mean rotation period of the star, $P_{\rm rot} = 4.522$~days,  is longer than the orbital period of the planet, $P_{\rm orb} = 1.743$~days. On the other hand, CoRoT-4 is  synchronized with a mean rotation period of $\sim 8.8$ days, very close to the orbital period of $9.202$~days, while CoRoT-6 is an example with an orbital period ($P_{\rm orb} = 8.889$~days) longer than the mean rotation period of the star ($P_{\rm rot}= 6.35$~days). We refer to the mean rotation period of the stars because they show surface differential rotation with a minimum amplitude of $\sim 3$ percent in the case of CoRoT-2 \citep{Silva-ValioLanza11}, {  $\sim 6$~percent for CoRoT-4 \citep{Lanzaetal09b}, } and $\sim 12$ percent for CoRoT-6 \citep{Lanzaetal11b}. {  These cases} are interesting also in view of the different methods applied to extract the signal associated with  SPMI according to the  ratio $P_{\rm rot}/P_{\rm orb}$. 

\subsection{CoRoT-2}

The light curve of CoRoT-2 has been extensively modelled by several researchers. \citet{Lanzaetal09a} applied a maximum entropy spot model to the out-of-transit light curve to derive the distribution of the covering factor of the active regions vs. longitude along successive rotations. A comparison of this approach with other spot modelling methods is provided by, e.g., \citet{Mosseretal09}, \citet{Frohlichetal09}, and \citet{Huberetal10}, while a detailed test with solar observations is presented by \citet{Lanzaetal07}. 
The modelling technique exploits the modulation of the visibility of the active regions by  the rotation of the star, so that a good phase coverage of each rotation is needed to derive a stable and reliable map. The advantage of CoRoT is the almost perfect time sampling (the duty cycle is virtually 100 percent for spot modelling), but the duration of the stellar rotation sets a minimum timescale for an accurate mapping that is $\sim 3$~days in our case, i.e., about 65 percent of a rotation. Morover, since the active regions evolve rapidly, it is not possible to obtain an adequate best fit for time intervals longer than $3-4$ days. 
In conclusion, the time resolution of the mapping of \citet{Lanzaetal09a} is longer than the orbital period of the planet making impossible a direct search for SPMI features in those maps.  

\begin{figure}
\includegraphics[width=5cm,angle=90]{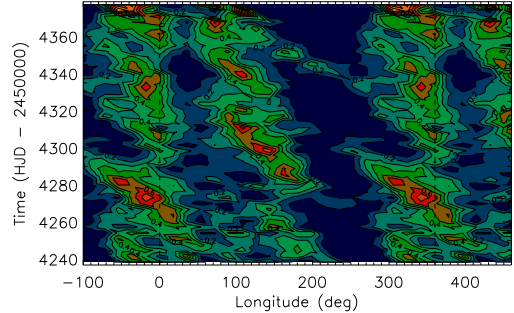} 
\caption{Maximum entropy spot model of CoRoT-2. The isocontours of the spot filling factor are reported vs. longitude and time. Different colours refer to different values of the filling factor with dark blue indicating the minimum (virtually no spots) and {  orange/yellow} the maximum. The longitude is measured in a reference frame rotating with the mean stellar rotation period of 4.5221 days and increases in the direction of the stellar rotation and the orbital motion of the planet {  which is prograde  \citep[cf. ][]{Bouchyetal08}}. Note that the longitude is repeated beyond the $0^{\circ}-360^{\circ}$ interval to help following the migration of the spots. Credit: \citet{Lanzaetal09a}, reproduced with permission, \copyright ~ESO.} 
\label{corot2_ff}
\end{figure}

\begin{figure}
\includegraphics[width=5cm,angle=90]{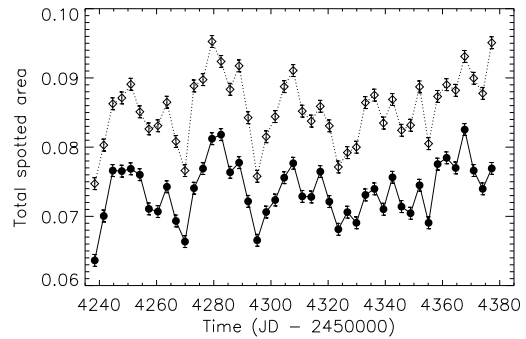} 
\caption{The total spotted area of CoRoT-2 vs. the time. The solid line refers to models  with dark spots only ($Q=0$), while the dotted line indicates the spot area in the models with a facular component with $Q=1.5$. 
Credit: \citet{Lanzaetal09a}, reproduced with permission, \copyright ~ESO.} 
\label{corot2_tot_area}
\end{figure}

The distribution of the active regions versus longitude and time on the surface of CoRoT-2 is plotted in  Fig.~\ref{corot2_ff} where we see two persistent active longitudes in which individual active regions appear, grow,  and decay with lifetimes up to $\sim 50-60$~days. The model by \citet{Lanzaetal09a} assumes that each active region  consists of dark spots and bright solar-like faculae with a facular-to-spotted area ratio indicated by $Q$. The models in Fig.~\ref{corot2_ff} have been computed for $Q=0.0$, i.e., without faculae. The variation of the total spotted area vs. time is plotted in Fig.~\ref{corot2_tot_area} and shows oscillations with a period of $29.6 \pm 4.0$~days, both in the model assuming only dark spots as well as in that with a facular component having $Q=1.5$. Note that for the Sun $Q=9$, while for more active stars, such as CoRoT-2, lower values of $Q$ are expected  \citep{Lanzaetal07}.


Two possible interpretations have been suggested for the oscillations of the total spotted area. The first considers them analogous to the oscillations of the total sunspot area observed close to some of the maxima of the eleven-year cycles in the Sun \citep{Oliveretal98,Zaqarashvilietal10} and displaying a periodicity of $150-160$~days. According to \citet{Lou00} and \citet{Zaqarashvilietal10}, Rossby-type magnetohydrodynamic waves, excited in the subphotospheric layers or at the interface between the convection zone and the radiative interior of the Sun, could account for such a short-term periodicity by modulating the emergence of magnetic flux. Since the period of the wave is proportional to the rotation period of the star, we expect a period of about one month in the case of CoRoT-2 that is a G-type star rotating five times faster than the Sun. 

An alternative explanation for the modulation of the spotted area in CoRoT-2 considers it as a signature of SPMI in the photosphere of the star {  because of  the close coincidence of the periodicity  with ten synodic periods of the planet with respect to the mean stellar rotation period}. The synodic period $P_{\rm syn}$  is the time interval between two successive passages of the same meridian across the subplanetary longitude in a rotating star. It is equal to $2.89$ days in the case of CoRoT-2, assuming a mean rotation period  $P_{\rm rot} = 4.5221$~days because it is given by: $P_{\rm syn}^{-1} = | P_{\rm rot}^{-1} - P_{\rm orb}^{-1} | $. According to this conjecture, the passage of the planet over an active region may trigger the emergence of magnetic flux tubes when their intensity is already close to the threshold for the onset of the buoyancy instability \citep{Acheson78, Acheson79}. Ten passages are to elapse  before the magnetic field intensity reaches again the threshold  needed for the planetary-induced perturbation to be effective. {  In Sect.~\ref{models}, a possible physical mechanism to account for such a perturbation will be introduced, based on the effects on the stellar dynamo induced by the reconnection in the stellar coronal field triggered by the planetary motion.}

Spots occulted by the planet during transit can be detected through the characteristic light bumps produced along the transit light curve when the planet's disc passes over them. {  The average spotted area of the occulted spots plotted vs. longitude shows six relative maxima, one of which is associated with the subplanetary longitude within $\pm 3^{\circ}$ \citep[cf. Fig.~4 of ][ where the subplanetary longitude is zero]{Silva-ValioLanza11}. }
\citet{Silva-ValioLanza11} investigated the variation of the total area {  of the occulted spots} and found a  periodicity of $17.7 \pm 2.3$~days. Although  very close to ten orbital periods of the planet, it is unlikely to be  an effect of the aliasing due to the periodic time sampling of the observations. In that case, we should observe other periodicities at, say, 5 or 15 orbital periods which is not the case \citep[see Fig.~10 in][]{Silva-ValioLanza11}. Moreover, computing the periodograms of 10~000 random permutations of the time series, 
a periodicity of 17.7 days occurs only in $\sim 3$ percent of the cases. 

The periodicity corresponds to $\sim 6$ synodic periods when the rotation period of the mean latitude of the strip occulted by the planet is considered, i.e., 4.48~days \citep{Silva-ValioLanza11}.   The occulted spots are a subset of all the spots present on the star and are located at a latitude between $\sim 5^{\circ}$ and $\sim 20^{\circ}$. On the other hand, the modulation of the light curve outside transit is produced by starspots at all latitudes. We expect that spots can be formed in a latitude range more extended than in the Sun because CoRoT-2 is at least 20 times more active than our star {  as measured by the amplitude of the optical flux modulation produced by the active regions} \citep[e.g. ][]{Strassmeier09}. Therefore, the periodicity of the starspot area of the band occulted during transits is diluted when we consider the starspot area  derived from the out-of-transit light curve. As a matter of fact, a peak corresponding to the second harmonic of the 17.7 day period is present in the periodogram of  the out-of-transit spotted area although its height is remarkably lower than that of the main peak at 29 days because of the effects of the dilution \citep[cf. Fig.~7 of ][]{Lanzaetal09a}. 

A possible interpretation of the 17.7 day periodicity that corresponds to $\sim 6$ synodic periods, is that the passage of the planet over the low-latitude active regions in the occulted band induces the same phenomenon as observed in the variation of the full-disc area. However, its cadence is shorter because the magnetic field is amplified more rapidly at low latitudes. This interpretation assumes that there are two latitudes where the conjectured  mechanism is operating, one in the occulted band, the other in the latitude range not occulted by the planet. The localization of the triggering in two latitude bands is required by the absence of intermediate periodicities in the area modulation. Moreover, these periodicities appear only when the total spotted area, i.e., integrated over the longitude, is considered. Therefore, the modulation is not a property of some specific active region, but of all the regions in a given latitude interval. 

Finally, \citet{Paganoetal09} considered the variance of the flux of CoRoT-2 in the CoRoT white passband vs. the orbital phase. The variance was calculated using the individual 32~s exposures binned vs. the orbital phase. Considering the first 75 days of the light curve when the jitter effects were minimal and a bin size of 0.05 in phase, they found that the variance of the flux varied regularly with the orbital phase reaching a maximum just immediately before the transit and a minimum around phase 0.4. This phenomenon could be related to an enhancement of the  activity in CoRoT-2  triggered by the passage of the planet, e.g., in the form of small white-light flares, {  maybe associated with the spots generally found at the subplanetary longitude during transits (see above).}

\subsection{CoRoT-4}

\begin{figure}
\includegraphics[width=5cm,angle=90]{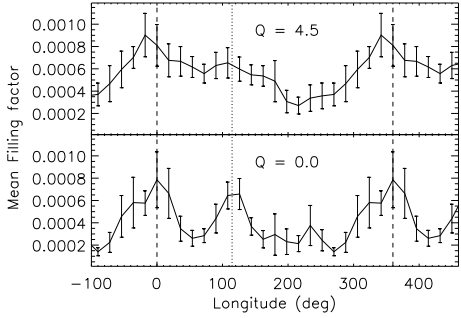} 
\caption{Averaged spot filling factor vs. longitude in CoRoT-4. The longitude is measured in a reference frame rotating with the orbital period of the planet and increases in the direction of the stellar rotation and the planetary motion, {  assumed to be prograde}. The top panel shows the spot model with a facular component in the active regions ($Q=4.5$), while the lower panel shows a model assuming only dark spots ($Q=0$). The vertical dashed lines  mark the longitudes $0^{\circ}$ and $360^{\circ}$ beyond which the plot is repeated, while the vertical dotted line marks the subplanetary longitude. Credit: \citet{Lanzaetal09b}, reproduced with permission, \copyright ~ESO.} 
\label{corot4_long}
\end{figure}

In the case of CoRoT-4, the mean rotation period of the star is  close to the orbital period of the planet. The spot modelling of \citet{Lanzaetal09b} shows a persistent active region located $\pm 50^{\circ}$ from  the subplanetary longitude, when a longitude reference frame rotating with the orbital period of the planet is assumed (see Fig.~\ref{corot4_long}). {  Note that the facular-to-spotted area ratio $Q$ adopted for the modelling affects somewhat the distribution of the spotted area vs. longitude. This happens because faculae are brighter when they are closer to the limb and thus they affect the distribution of the dark spots to reproduce the observed light modulation \citep[cf., e.g., ][]{Lanzaetal07}. Nevertheless, the presence of spots within $\pm 40^{\circ}$ from the subplanetary longitude is confirmed by the modelling of the light bumps associated with the spots occulted during the transits (Silva-Valio \& Lanza 2011, in prep.).}

The star shows a  differential rotation of at least $\sim 6$~percent which is not unusual for an F-type star. In spite of that, the active longitude associated with the planet has persisted for {  at least} $\sim 60$~days, i.e., the duration of the CoRoT observations. Unfortunately, this time span is too short to exclude that the coincidence happened by chance, but the phenomenology is reminiscent of what \citet{Walkeretal08} observed in $\tau$~Boo, another F-type star in almost synchronous rotation with the orbit of its hot Jupiter. In that case, in spite of a remarkable differential rotation \citep[$\sim 18$~percent; see ][]{Catalaetal07}, a persistent active region was found leading the planet by $\approx 80^{\circ}$  with a lifetime of at least $\approx 1400$ days. {  \citet{Mosseretal09} found that photospheric spots have typical lifetimes not exceeding one week in F-type stars which makes such a long persistence remarkable. The possibility that the feature is associated with the tidal interaction is ruled out by the lack of a similar feature on the opposite hemisphere of the star, i.e., associated with the other tidal bulge.} 

\subsection{CoRoT-6}

\begin{figure}
\centerline{\includegraphics[width=5cm,angle=90]{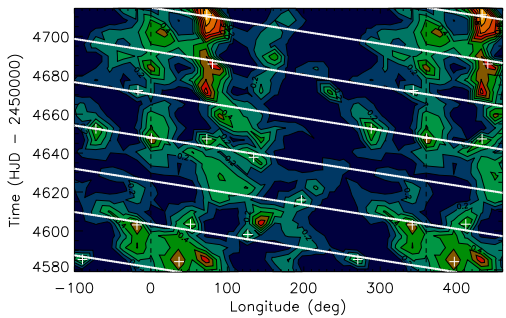}} 
\caption{Maximum entropy spot model of CoRoT-6 with $Q=1.5$. The isocontours indicate different values of the starspot filling factor with the same color coding of Fig.~\ref{corot2_ff}. The adopted longitude reference frame is rotating with the mean rotation period of the star ($P_{\rm rot}=6.35$~days). The longitude increases in the direction of the stellar rotation and the orbital motion of the planet (which is prograde), and is repeated beyond the $0^{\circ}-360^{\circ}$ interval to help following the migration of the spots. The white crosses mark the relative maxima of the spot filling factor in the active regions, while the white straight lines indicate a longitude lagging the subplanetary longitude by $200^{\circ}$. Credit: \citet{Lanzaetal11b}, reproduced with permission, \copyright ~ESO.} 
\label{corot6_ff}
\end{figure}

A maximum entropy spot modelling of CoRoT-6 has been presented by \citet{Lanzaetal11b} and shows an interesting phenomenology. Now the rotation period ($P_{\rm rot}=6.35$ days) is shorter than the orbital period of the planet 
($P_{\rm orb}= 8.886$~days), giving us the possibility to resolve the spot evolution during the orbital as well as  the synodic periods  ($P_{\rm syn}=22.25$~days). Relative maxima of the spotted area occur in {  about ten}  active regions  when a longitude lagging the planet by $\sim 200^{\circ}$ passes by the regions themselves, as shown in Fig.~\ref{corot6_ff}. The probability of a chance association has been estimated by \citet{Lanzaetal11b} to be lower than 1~percent. Again, the passage of the planet over the stellar active regions seems to be associated with a triggering of new magnetic flux emergence, although there is an average time lag of $\sim 3.5$~days between the passage of the planet and the maximum spotted area in the active regions. The total spotted area, {  as derived from the out-of-transit light curve,} does not show any obvious periodicity, possibly because the synodic period is comparable or longer than  the  lifetime of the active regions, so the planet-induced modulation is hidden by the larger variations associated with the  intrinsic growth and decay of the active regions. {  On the other hand, a preliminary analysis of the variation of the total spotted area occulted during transits shows a modulation with a period of $22.5-24$~d, close to the synodic period of the planet (Silva-Valio \& Lanza 2011, in prep.). }

{  \subsection{Comparison with some preliminary Kepler results}

The preliminary results coming from the analysis of the light curves of the Kepler space telescope \citep[][]{Boruckietal09} show two interesting cases of possible star-planet interaction in the photosphere. \citet{Sanchis-OjedaWinn11} analysed the transit light curves of HAT-P-11, a K4V star, confirming that the system is strongly misaligned as already known from the observations of the Rossiter-McLaughlin effect. The interesting feature is that the spot occultations by the planet occur only at two fixed phases during the 26 observed transits. This is explained by assuming that the planet transits over the starspots located in two low-latitude activity belts on a star viewed almost equator-on, similar to what we observe in the Sun where sunspots appear in two active latitudes symmetric with respect to the equator that migrate towards the equator as the cycle progresses. Another possibility is that the star is viewed almost pole-on, in which case the transit chord crosses twice the same high-latitude spot belt \citep[cf. Fig.~7 in ][]{Sanchis-OjedaWinn11}. The intriguing feature is that starspots are observed at the first intersection between the transit chord and the active belts in 13 out of 26 transits and at the second intersection in 13 out of 26 transits. 
In other words, in the 50 percent of the transits we observe a starspot not too far from the subplanetary longitude. 
However, since the star is about five times more active than the Sun at the maximum of the eleven-year cycle, as measured by the amplitude of the optical light curve, this close association between the planet and the occulted spots could have occurred simply by chance. 

A more impressive case is that of Kepler-17, a G-type star quite similar to the Sun, except for its rotation period of $11.9$~d, and accompanied by a hot Jupiter with an orbital period of 1.486~d \citep{Desertetal11}. The rotation period of the spots occulted during the $\sim 180$ observed transits is just eight times the planet orbital period producing a particular stroboscopic effect in their visibility. Five occulted active regions show a lifetime of at least $\sim 100$~d and reach their maximum filling factor when they cross the subplanetary longitude corresponding to phase 0 in the right hand plots in Fig.~11 of \citet{Desertetal11}. This remarkable association between the planet and the occulted starspots is confirmed by the average transit light curve  which shows a clear bump centred at phase 0 \citep[cf. Fig.~2 in ][]{Desertetal11}. In view of the large number of observed transits covering a time interval of 16.7 months, such an association  is highly unlikely to have occurred by chance. Moreover, the phenomenology is remarkably similar to that of CoRoT-6, in spite of the differences in the stellar and planetary parameters.  

}

\section{Theoretical models}
\label{models}

\subsection{Chromospheric and coronal features of SPMI}
\label{chromo_features}

Several theoretical models have been proposed to account for specific features of SPMI as observed in the chromosphere and possibly in the corona, but a comprehensive 
theory is still lacking. To explain the presence of chromospheric hot spots rotating in phase with the planet, the first idea was that of considering a scaled version of the unipolar induction model proposed for the Jupiter-Io system. Observations in the UV has revealed two bright spots in  the Northern and the Southern hemispheres  of Jupiter, respectively, located at the footpoints of the  loops connecting the magnetic poles of Jupiter with Io. A flux of Alfven waves is excited by the motion of Io across the magnetic field lines of  Jupiter's large-scale dipole  field and the energy of the waves is conducted down to the loop footpoints producing the emission \citep[see, e.g., ][ and references therein]{Zarka07}. This mechanism can work even if Io has no intrinsic magnetic field because it is sufficient its orbital motion across the Jupiter's field lines and some surface electric conductivity to  excite the  Alfven waves. 

A limitation of the model in the stellar case is the low power that can be conveyed onto the chromosphere, estimated of the order of $10^{16}-10^{17}$~W. 
The observed phase lag between the planet and the hot spot is also difficult to explain with a potential dipole magnetic field like that of Jupiter.  \citet{McIvoretal06} suggested that the axis of the stellar dipole is tilted with respect to the orbital angular momentum of the planet, but even with this hypothesis it is not possible to account for a phase lag of $\sim 160^{\circ}$ as observed in the case of $\upsilon$~And. On the other hand, the model by \citet{Preusseetal06}, recently confirmed by the numerical simulations of \citet{Koppetal11}, can explain a large phase lag. It is based on packets of Alfven waves that are excited by the orbital motion of the planet, even in the absence of an intrinsic planetary field, and then propagate towards the star along characteristics making some angle  with the magnetic field lines. They can reach the surface provided that the velocity of the stellar wind at the distance of the planet is subalfvenic, that  is the case when the planet is sufficiently close to the star. The main limitation  is the low energy flux at the stellar surface that makes almost impossible to account for an emitted power as high as $10^{20}-10^{21}$~W. {  Note that the same order of magnitude is required to explain the enhancement of the X-ray emission claimed by \citet{Kashyapetal08}.} 	

To explain the emitted power, \citet{Lanza09} considered the possibility that the interaction between the planetary magnetic field and the stellar coronal field may trigger the release of the energy already stored in the coronal loops. The reconnection process continuously produced by the planetary field in the outer corona tends to reduce the magnetic energy of the coronal field. Since the minimum energy state for a given total magnetic helicity is a linear force-free field, the geometry of the field lines can be described analytically given the boundary conditions at the stellar surface. An interesting property of linear force-free fields is that of being twisted, thus providing a natural explanation for the phase lag between the planet and the chromospheric hot spot located close to the footpoints of the magnetic field lines. This property has been explored in detail by \citet{Lanza08}. In combination with the triggering of magnetic energy release suggested by \citet{Lanza09}, force-free fields could account for the emitted power of the hot spots as well as for some coronal emission enhancement. Depending on the total magnetic helicity and the surface boundary conditions, the magnetic field lines intersecting the planetary magnetosphere may close up before reaching the surface of the star which may imply that no magnetic connection exists between the planet and the chromosphere \citep[see ][]{Lanza09}. Such configurations would give rise to an "off" state of the SPMI in the terminology of \citet{Shkolniketal08}. When the boundary conditions and/or the total helicity change, a field configuration that connects the planet with the star can be resumed leading to a switching on of the SPMI features in the chromosphere. Another interesting property of linear force-free fields is the predominance of closed field lines that reduce the loss of angular momentum in the stellar wind and the consequent braking of the stellar rotation. This may have important consequences for the estimate of the age of the stars accompanied by hot Jupiters through the method of gyrochronology \citep{Lanza10}.

Full magnetohydrodynamic simulations, such as those of \citet{Ipetal04},  \citet{Cohenetal09a}, \citet{Cohenetal09b}, \citet{Cohenetal10}, \citet{Cohenetal11a}, and \citet{Cohenetal11b}, confirm several results of the analytical force-free models by \citet{Lanza08,Lanza09,Lanza10} and show how a stellar corona tends to be confined and heated by the interaction with the magnetosphere of a hot Jupiter. Hot coronal regions can account for an increased emission in the X-rays and the reduced efficiency of the acceleration of the stellar wind. Although the magnetohydrodynamic regimes accessible to numerical simulations are many orders of magnitude  far from  real systems, these studies provide a wealth of information on what is to be expected in real star-planet interaction and can be specialized to the case of particular systems \citep[cf., e.g., ][]{Cohenetal11a}.

Another approach to the star-planet interaction has considered individual coronal loops reconnecting  with the planetary magnetosphere. According to the phase of the stellar activity cycle, the height of the tops of the loops can vary leading in some cases to a strong interaction, modulated with the orbital period of the planet. On the other hand, during phases when the loops are not so tall the interaction with the planetary field is reduced and we predominantly see the modulation by the stellar rotation \citep{CranmerSaar07}. 
{  Thus this mechanism can naturally account for the on/off transitions in the SPMI reported by \citet{Shkolniketal05,Shkolniketal08}.}
Other studies have addressed the  transport of  energy from the reconnection site, presumably in the corona, to the chromosphere through beams of energetic particles that produce a localized heating when they impact onto the chromosphere \citep{GuSuzuki09}. 

Finally, we should mention the models developed to predict the radio emission from the systems hosting hot Jupiters. Since those planets are much closer than Jupiter to the Sun, the stellar wind speed is likely to be subalfvenic. However, a remarkable interaction {  between the wind and the planetary magnetosphere} is predicted, and the emitted radio fluxes should be of the order of $10^{5}$ those of Jupiter. The peak of the emitted spectrum should fall at frequencies of a few or several tens of MHz depending on the strength of the planetary magnetic field \citep[see, e.g., ][ and references therein]{Stevens05,Zarka07,JardineCameron08,Lanza09,HessZarka11}. A significant flux at microwave  frequencies is expected if high-energy particles spiral along the kG magnetic fields of the spots in the stellar photosphere, but to attribute the  emission to SPMI it is necessary to detect its modulation with the orbital or the synodic period. On the other hand, radio emission in the MHz range is characteristic of the planetary fields and its detection would   provide  information on the strength of the planetary magnetic fields.  

\subsection{Accounting for the energy budget}
\label{energy_bud}

Most of the models considered above dealt with  analogies or differences between the solar-planet and the star-planet interactions, or focussed on specific issues such as the phase lag between the chromospheric hot spot and the planet. Nevertheless, only some of them considered the problem of dissipating a power of the order of $10^{20}-10^{21}$~W to account for  the energy irradiated by the chromospheric hot spots of \citet{Shkolniketal05}. In their MHD simulations, \citet{Ipetal04} were able to obtain powers of the order of $3 \times 10^{19}$~W adopting a magnetic field of $0.1$~G and a size of the interaction area of 5~Jupiter radii at the magnetopause between the planetary field and the stellar coronal field with an orbital velocity of the planet of $280$~km~s$^{-1}$. However, typical surface fields in HD~179949 are of the order of $10$~G at most and even in the case of the remarkably active host HD~189733 they do not exceed $\sim 40$~G \citep{Moutouetal07}. Since the magnetic field strength decreases at least as $(r/R)^{-3}$, where $r$ is the distance from the centre of the star and $R$ its radius, we expect  field strengths  $B \simeq 0.02-0.08$~G at a distance of  $8R$, typical of most of the systems with hints of SPMI. Since the dissipated power is proportional to $B^{2}$, the field strength is too low  to account for the energy radiated in the chromospheric hot spots. A power at least two or three orders of magnitude larger is required. 

\begin{figure*}
\includegraphics[width=16cm]{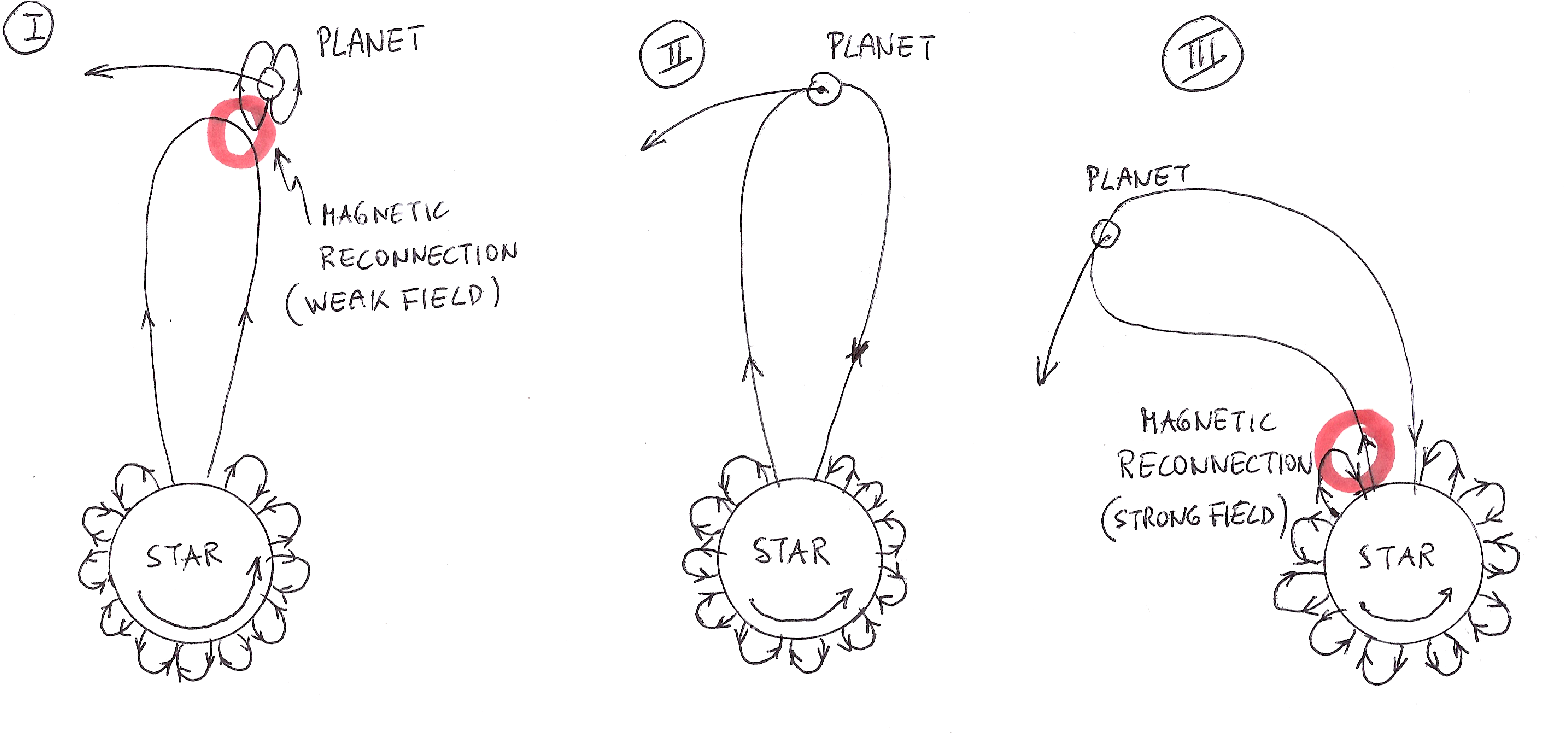} 
\caption{A sketch of the three main phases of the conjectured star-planet interaction in the outer stellar corona. {  In the first stage (left panel), the magnetic reconnection connects a tall coronal loop with the planetary magnetosphere in a region of weak field intensity ($B \sim 0.01$~G); in the second stage (middle panel), the interconnecting loop is stressed and bent by the orbital motion of the planet that is not synchronized with the stellar rotation; finally, in the third stage (right panel), the energy stored in the interconnecting loop is released when it reconnects with low magnetic loops close to the star in a region of relatively strong ($B \sim 1$~G) field. }} 
\label{coronal_sketch}
\end{figure*}

\citet{Lanza09} proposed that the energy needed to explain the hot spot  comes from the whole magnetic structure connecting the planet with the star and is released at a closer distance than the planet orbital radius. His considerations were based on models borrowed from solar physics to explain the budget of solar flares and {  considered the fundamental role played by magnetic helicity in the evolution of the coronal field.} Here, we  introduce some intuitive ideas for such a kind of models by means of a simple cartoon (see Fig.~\ref{coronal_sketch}). We consider a late-type star with a hot Jupiter whose orbital motion is not synchronous with stellar rotation. At the beginning of the process, a long coronal loop reconnects with the planetary magnetosphere releasing a modest amount of energy, with a characteristic dissipated power of the order of $10^{17}$~W for a time scale of the order of $L/v_{\rm A}$, where $L$ is the size of the interaction region and $v_{\rm A}$ the Alfven velocity. Once the reconnection process is over, the relative orbital motion of the planet with respect to the loop footpoints produces a bending of the loop on a timescale of the order of the day. The stress of the field accumulates energy into the loop and the field is no longer potential as indicated by the curvature of the field lines in the outer corona. This energy can be released closer to the star when the curved field lines  interact with shorter loops, as in the final sketch of the cartoon. Since  the interaction site is closer to the star, the magnetic field intensity is remarkably larger there, while the relative velocity field, that scales roughly as $(r/R)$, is still of the order of several tens of km~s$^{-1}$. Considering a field strength $B= 1$~G at the interaction site, a typical size of the interacting region of $L = R$, and a relative velocity $v = 30$ km~s$^{-1}$, we have a released power of: 
\begin{equation}
P_{\rm diss} \simeq \gamma \frac{\pi}{\mu} B^2 L^2 v \simeq 3.5 \times 10^{20} \gamma \mbox{~W},
\end{equation}
where $0 < \gamma < 1$ is a factor depending on the relative angle between the reconnecting magnetic field lines at the interaction site and $\mu$ is the magnetic permeability of the plasma \citep[cf. Eq.~(8) of ][]{Lanza09}. For oppositely directed fields, we get the maximum efficiency with $\gamma \simeq 1$ and we have a power sufficient to explain the energy radiated by chromospheric hot spots. Note, however, that ordinary stellar flares release a  comparable power when the relative velocity between magnetic loops is of the same order. The effect of the planet is that of inducing a sequence of flares when an interconnecting loop is stressed again by the relative motion of the planet. Since most of the energy is released close to the star, the chromospheric heating does not require any ad hoc process to convey the energy from the planetary orbit to the chromosphere. The only requirement is the presence of an interconnecting loop whose field lines are stressed by the relative motion of the planet without producing an instability that destroys the loop itself before it have had time to accumulate enough energy. {  An important factor is the transverse size of the interconnecting loop, that we assume of the order of the radius of the star. It depends also on the intensity of the magnetic field of the planet because a stronger field produces a more extended magnetosphere \citep[cf. Eq.~(7) in ][]{Lanza09} that increases the probability of reconnection with stellar loops and thus the transverse size of the interconnecting loop. The energy extracted from the orbital motion and stored in the magnetic interconnecting loop should in principle lead to a slow orbital decay of the planet. Since the total mechanical energy of the orbital motion is $GM_{\rm s}m_{\rm p}/2a$, where $G$ is the gravitation constant, $M_{\rm s}$ the mass of the star, and $m_{\rm p}$ that of the planet, a typical orbital energy of $10^{36}-10^{37}$~J is obtained for a hot Jupiter. With the dissipated powers considered above, this implies an orbital decay timescale of at least several Gyr. Therefore, the tidal interaction is likely to be more effective for the evolution of the orbit of the planet \citep[cf., e.g., ][]{GoupilZahn08}.  }

This promising model for the magnetic star-planet interaction, here sketched with simple order-of-magni\-tude and qualitative considerations, will be presented in more detail in a forthcoming study (Lanza 2011, in prep.).  Note that in systems close to synchronizations, such as $\tau$~Boo or CoRoT-4, the relative velocity $v$ is expected to be remarkably smaller than in the case discussed above. However, surface differential rotation may still provide sufficient relative velocity for the onset of the process, although with a reduced efficiency. 

%

\subsection{A conjecture on the perturbation of the stellar dynamo by the planet}
\label{conjecture}

{  The photospheric phenomena presented in the case of the CoRoT and Kepler targets in Sect.~\ref{phot_SPMI_corot}  or  in \citet{Walkeretal08}
for the case of $\tau$~Boo cannot be explained with the models discussed so far because they consider only the interaction and the associated phenomena in the outer atmosphere. Nevertheless, as we shall see in  this section, there could be a connection between the perturbation produced by the planet on the magnetic field of the corona and the operation of the stellar dynamo responsible for the amplification and modulation of the stellar field appearing in the photosphere.} 

The emergence of new magnetic flux in the form of spots or faculae requires some mechanism to trigger the buoyancy instability of the magnetic field when the planet passes over an active region. The first conjecture is that the tidal deformation of the stellar field may promote its instability if the field is already close to the threshold value \citep{Acheson79}. \citet{HolzwarthSchuessler03a,HolzwarthSchuessler03b} have studied tidal effects on the stability and evolution of magnetic flux tubes in stellar interiors in the case of close binary systems. The maximum tidal perturbation occurs for flux tubes stored in the subphotospheric layers. \citet{Brandenburg05} has recently discussed the possibility that the solar dynamo has a source region located in the subphotospheric layers of the convection zone, although other authors, e.g., \citet{Dikpatietal02}, pointed out several difficulties for the dynamo to operate there. Nevertheless, the possibility of a flux storage not far from the surface  cannot be excluded in a magnetically active star. 

The main difficulty with this model is that tidal effects are virtually negligible for CoRoT-4 and CoroT-6 since the ratio of the semimajor axis of the orbit to the stellar radius $a/R \sim 17-19$ in those systems and the tidal deformation scales as $(a/R)^{-3}$. Moreover, two tidal bulges are expected, separated by $180^{\circ}$ in longitude, but only one active longitude is associated with the planet in those stars. Therefore, it is unlikely that a tidally-based mechanism can account for the observations.

An alternative mechanism has been suggested by \citet{Lanza08} in the appendix to his paper. It assumes that the stellar dynamo is somehow perturbed by the planet. Two main effects are responsible for the magnetic field generation and modulation in late-type stars: differential rotation and the $\alpha$ effect, the latter related to the mean helicity of the turbulent motions. Differential rotation stresses poloidal fields directed along the meridians and produces a toroidal field component, i.e., parallel to the equator, that gives rise to the sunspots after emerging as discrete flux tubes. The $\alpha$ effect is responsible for the re-generation of the poloidal field from the toroidal field allowing the dynamo to operate. As discussed in, e.g., Ch.~9 of \citet{BrandenburgSubramanian05}, also the mean current helicity associated with small-scale magnetic fields plays a role in generating the $\alpha$ effect. The expression for the $\alpha$ effect that takes into account  the current helicity is:
\begin{equation}
\alpha  \propto (\langle {\vec v} \cdot \nabla \times {\vec v} \rangle - \langle {\vec j} \cdot {\vec b} \rangle) \tau,
\end{equation}
where $\vec v$ is the turbulent velocity field, $\vec b$ the turbulent magnetic field, ${\vec j} \equiv \nabla \times {\vec b}$ the turbulent current density, the angular brackets denote an ensemble average, and $\tau$ is a characteristic correlation time between the turbulent fields. The mean current helicity is related to the evolution of the total helicity of the magnetic field of the star by the law of helicity conservation \citep[see Chs. 3 and 9 of][]{BrandenburgSubramanian05}. Specifically, the variation of the  magnetic helicity of the star is given by:
\begin{equation}
\frac{d}{dt} \int_{V} {\vec A} \cdot {\vec B} \; dV = -2 \eta \int_{V} {\vec J} \cdot {\vec B} \; dV,
\end{equation}
where $\vec A$ is the vector potential of the magnetic field ${\vec B} \equiv \nabla \times {\vec A}$, $V$  the volume of the star, $\eta$  the magnetic diffusivity of the plasma assumed to be uniform, and $\vec J \equiv \nabla \times {\vec B}$  the current density. For an ideal plasma $\eta =0$ and the integral in the l.h.s., i.e., the magnetic helicity, is conserved. In the presence of dissipation, the rate of change of the magnetic helicity
depends on the diffusivity and the integral in the r.h.s., i.e., the current helicity. If we assume that the timescale for the variation of the magnetic helicity is much longer than the timescale for the variations of the currents, i.e., the dynamo operates in an almost stationary regime, then:
\begin{equation}
\int_{V} {\vec J} \cdot {\vec B} \; dV \simeq 0.
\label{curr_hel}
\end{equation}
Considering the current density and the magnetic field as the sum of their average and fluctuating components, viz. $\vec J = \langle {\vec J} \rangle + {\vec j}$ and $\vec B = \langle {\vec B} \rangle + {\vec b}$,  Eq.~(\ref{curr_hel}) and the Reynolds  rules  give:
\begin{equation}
\int_{V} \langle {\vec J} \rangle \cdot \langle {\vec B} \rangle \; dV  \simeq - \int_{V} \langle  {\vec j}  \cdot  {\vec b} \rangle  \; dV.
\label{eq_int} 
\end{equation}
Equation~(\ref{eq_int}) implies that the loss of the mean current helicity occurring at the reconnection sites in the corona, that changes the integral in the  l.h.s., produces a variation of the small-scale helicity on the r.h.s. This in turn perturbs the $\alpha$ effect. 
Specifically, the reconnection events triggered by the planet in the corona (see Sect.~\ref{energy_bud}) may  induce a non-axisymmetric  component of the $\alpha$ effect  modulated in phase with the passage of the planet across the large scale loops. This phase-dependent $\alpha$ effect can produce a non-axisymmetric magnetic field that emerges at the surface where the intensity of the unperturbed magnetic field is already close to the threshold for the onset of the buoyancy instability. 

In conclusion, the perturbation of the $\alpha$ effect at the longitude where magnetic reconnection is more effective,  owing to the interaction with the planet, leads to a modulation of the field emergence with the orbital period of the planet or its synodic period. The timescale for such a modulation depends on the intensity of the background field and its amplification timescale that are ruled by the unperturbed stellar dynamo. In the case of CoRoT-2, we  conjecture that two interconnecting loops are responsible for the observed modulation. The one with the footpoints in the latitude range occulted by the planet is located in a zone of higher dynamo efficiency leading to a typical timescale of $6$ synodic periods for the magnetic flux modulation. On the other hand, the one located at higher latitudes, where the dynamo is less efficient, requires $ 10$ synodic periods between successive emergence episodes because the dynamo takes longer to bring the field close to the threshold. 

\section{Discussion and conclusions}

We have reviewed the current evidence of star-planet magnetic interaction in stellar atmospheres {  focussing on the photospheres}. Unfortunately, the current results {  neither point definitely to a single type of phenomena nor can be explained by a single model,}  so further dedicated observations are needed to assess the reality and clarify the properties of the related phenomena. {  Nevertheless, one common aspect of the different phenomenologies appears to be the presence of starspots whose formation is triggered by the passage of the planet over stellar active longitudes, and that are often localized close to the subplanetary longitude. }

Space-borne long-term photometry such as that made possible by MOST, CoRoT, Kepler (and later PLATO)  provides unique data sets to investigate the presence of photospheric starspots related with a  hot Jupiter, as we discussed in the cases of CoRoT-2, CoRoT-4, CoRoT-6, and Kepler targets. Ground-based searches should concentrate on the monitoring of chromospheric proxies, such as Ca~II~H~\&~K line core emission, to obtain a sufficient statistics on the properties of chromospheric hot spots. {  Those observations provide basic constraints to the magnetic interaction in the outer atmospheric domains that may induce perturbations of the stellar dynamo leading to the formation of starspots associated with the planet (cf. Sect.~\ref{conjecture}). }

Theoretical models should clarify the mechanisms for the interaction and this requires a map of the coronal field structure. As in the case of the Sun, it is possible to extrapolate the field components measured at the photosphere. Since the high-order multipoles of the photospheric field decay rapidly with the distance, only the low-order components, say dipole and quadrupole, are relevant at the distance of the planet. This means that the present photospheric field reconstructions based on spectropolarimetric techniques are  adequate to extrapolate the field at the distance of the planetary magnetosphere \citep[e.g., ][]{Moutouetal07,Faresetal10}. The simultaneous observations of a chromospheric hot spot and  the photospheric field to extrapolate the coronal configuration, will provide unique information to understand the interaction \citep[see ][ for details]{Lanza09}. 

The possibility that the planet somehow triggers the emergence of new magnetic flux from the interior, appearing as starspots or faculae in the photosphere, cannot be excluded and seems to be supported by the observations of the CoRoT {  and Kepler} targets discussed above. If confirmed in other systems, this will open a new view on stellar dynamos, planetary magnetic fields and star-planet interaction making this area of research of the highest interest for the advancement of stellar and planetary astrophysics. 

\acknowledgements
This paper has been developed from a review originally presented at the 
2nd CoRoT Symposium. The author wishes to thank the SOC for their kind 
invitation to attend the Conference and review the status
star-planet interaction studies. The author also wishes to thank
Prof. M. Deluil, Prof. J. Linsky, Dr. C. Moutou, and
Dr. A. S. Bonomo for interesting discussions on several
aspects of the star-planet magnetic interaction. Many
thanks also to the Editor in Chief of Astrophysics and
Space Science, Prof. M. A. Dopita, for his kind invitation
to submit this paper as an invited review.
Last but not least, the insightful comments by an anonymous 
referee proved to be a great help in improving this work.

\end{document}